\documentclass[aps,prl,preprint,tightenlines,superscriptaddress,showpacs,
byrevtex,floatfix]{revtex4}
\usepackage{dcolumn}  

\usepackage{epsfig}
\newcommand{\mpp}{M_{p\bar{p}}}
\newcommand{\mb}{{M_{\rm bc}}}
\newcommand{\de}{{\Delta{E}}}

\newcommand{\pp}{{p\bar{p}}}
\newcommand{\LL}{{\Lambda\bar{\Lambda}}}
\newcommand{\LLK}{{\Lambda\bar{\Lambda}K^+}}
\newcommand{\ppk}{{p\bar{p}K^+}}

\begin{document}
\preprint{\vbox{ \hbox{   }
                  \hbox{BELLE-CONF-0534}
}}
\title{ \quad\\[0.5cm]
\boldmath Study of Charmonium Decays into Baryon-Antibaryon Pairs}



\affiliation{Aomori University, Aomori}
\affiliation{Budker Institute of Nuclear Physics, Novosibirsk}
\affiliation{Chiba University, Chiba}
\affiliation{Chonnam National University, Kwangju}
\affiliation{University of Cincinnati, Cincinnati, Ohio 45221}
\affiliation{University of Frankfurt, Frankfurt}
\affiliation{Gyeongsang National University, Chinju}
\affiliation{University of Hawaii, Honolulu, Hawaii 96822}
\affiliation{High Energy Accelerator Research Organization (KEK), Tsukuba}
\affiliation{Hiroshima Institute of Technology, Hiroshima}
\affiliation{Institute of High Energy Physics, Chinese Academy of Sciences, Beijing}
\affiliation{Institute of High Energy Physics, Vienna}
\affiliation{Institute for Theoretical and Experimental Physics, Moscow}
\affiliation{J. Stefan Institute, Ljubljana}
\affiliation{Kanagawa University, Yokohama}
\affiliation{Korea University, Seoul}
\affiliation{Kyoto University, Kyoto}
\affiliation{Kyungpook National University, Taegu}
\affiliation{Swiss Federal Institute of Technology of Lausanne, EPFL, Lausanne}
\affiliation{University of Ljubljana, Ljubljana}
\affiliation{University of Maribor, Maribor}
\affiliation{University of Melbourne, Victoria}
\affiliation{Nagoya University, Nagoya}
\affiliation{Nara Women's University, Nara}
\affiliation{National Central University, Chung-li}
\affiliation{National Kaohsiung Normal University, Kaohsiung}
\affiliation{National United University, Miao Li}
\affiliation{Department of Physics, National Taiwan University, Taipei}
\affiliation{H. Niewodniczanski Institute of Nuclear Physics, Krakow}
\affiliation{Nippon Dental University, Niigata}
\affiliation{Niigata University, Niigata}
\affiliation{Nova Gorica Polytechnic, Nova Gorica}
\affiliation{Osaka City University, Osaka}
\affiliation{Osaka University, Osaka}
\affiliation{Panjab University, Chandigarh}
\affiliation{Peking University, Beijing}
\affiliation{Princeton University, Princeton, New Jersey 08544}
\affiliation{RIKEN BNL Research Center, Upton, New York 11973}
\affiliation{Saga University, Saga}
\affiliation{University of Science and Technology of China, Hefei}
\affiliation{Seoul National University, Seoul}
\affiliation{Shinshu University, Nagano}
\affiliation{Sungkyunkwan University, Suwon}
\affiliation{University of Sydney, Sydney NSW}
\affiliation{Tata Institute of Fundamental Research, Bombay}
\affiliation{Toho University, Funabashi}
\affiliation{Tohoku Gakuin University, Tagajo}
\affiliation{Tohoku University, Sendai}
\affiliation{Department of Physics, University of Tokyo, Tokyo}
\affiliation{Tokyo Institute of Technology, Tokyo}
\affiliation{Tokyo Metropolitan University, Tokyo}
\affiliation{Tokyo University of Agriculture and Technology, Tokyo}
\affiliation{Toyama National College of Maritime Technology, Toyama}
\affiliation{University of Tsukuba, Tsukuba}
\affiliation{Utkal University, Bhubaneswer}
\affiliation{Virginia Polytechnic Institute and State University, Blacksburg, Virginia 24061}
\affiliation{Yonsei University, Seoul}
  \author{K.~Abe}\affiliation{High Energy Accelerator Research Organization (KEK), Tsukuba} 
  \author{K.~Abe}\affiliation{Tohoku Gakuin University, Tagajo} 
  \author{I.~Adachi}\affiliation{High Energy Accelerator Research Organization (KEK), Tsukuba} 
  \author{H.~Aihara}\affiliation{Department of Physics, University of Tokyo, Tokyo} 
  \author{K.~Aoki}\affiliation{Nagoya University, Nagoya} 
  \author{K.~Arinstein}\affiliation{Budker Institute of Nuclear Physics, Novosibirsk} 
  \author{Y.~Asano}\affiliation{University of Tsukuba, Tsukuba} 
  \author{T.~Aso}\affiliation{Toyama National College of Maritime Technology, Toyama} 
  \author{V.~Aulchenko}\affiliation{Budker Institute of Nuclear Physics, Novosibirsk} 
  \author{T.~Aushev}\affiliation{Institute for Theoretical and Experimental Physics, Moscow} 
% ...
\collaboration{The Belle Collaboration}


\begin{abstract}
 
We study the baryonic charmonium decays of $B$ mesons, $B^+ \to \eta_c K^+$ and
$B^+ \to J/\psi K^+$, where $\eta_c$ and $J/\psi$ subsequently
decay into a $p\bar p$ or $\Lambda\bar\Lambda$ pair. 
The charmonium produced in the above $B$ meson decays is fully polarized. The polar angular distributions
of the baryon-antibaryon pairs are presented, along with fit results to a
$1 + \alpha_B\cos^2\theta$ parametrization. Comparisons are made with the
results from $e^+e^- \to J/\psi$ formation experiments. We also report the first
observation of $\eta_c \to \LL$. The measured branching fraction is
${\mathcal B}(\eta_c \to \LL) = ( 0.87^{+0.24}_{-0.21} \pm 0.14 \pm 0.27)
\times 10^{-3}$.
This study is based on a $357~fb^{-1}$ data sample recorded
on the $\Upsilon({\rm 4S})$ resonance with the Belle detector at
the KEKB asymmetric-energy $e^+e^-$
collider.

\end{abstract}
\maketitle
\tighten
{\renewcommand{\thefootnote}{\fnsymbol{footnote}}
\setcounter{footnote}{0}

There have been many reported observations of baryonic three-body $B$ 
decays in recent years~\cite{ppk,plpi,pph,LLK}.
An interesting feature of these observations is the presence of a
peak near threshold
in the mass spectra of the baryon-antibaryon pair.
Studies show that
these enhancements are not likely to be resonance states,
as the baryon angular distributions are not symmetric in their respective
helicity frames~\cite{polar}.
Other visible structures in the
mass spectra arise from charmonium decays. 
It is natural to compare the baryon angular distributions from charmonium decays
with those in the region of the threshold enhancement. 
There is a particular interest in $J/\psi \to \pp$, where the 
proton angular distribution has been studied by
DASP~\cite{DASP}, DM2~\cite{DM2}, MarkI~\cite{MarkI},
MarkII~\cite{MarkII} and BES~\cite{BES}~\cite{Bai:1998fu}.
$J/\psi$ mesons produced in $e^+e^-\to J/\psi$ are transversely polarized.
Accordingly, the baryon angular distribution 
can be parameterized as $\sim 1+\alpha \cos^2\theta$, 
where $\theta$ is the baryon polar angle
in the  $J/\psi$ helicity frame. Many theoretical 
predictions~\cite{theory} exist for the value of $\alpha$. The current world average value of $\alpha$, 
obtained with above measuremens from $J/\psi \to p\bar p$ decays, is $0.66 \pm 0.05$.

The study of two-body baryonic decays of charmonia at a $B$-factory
has several different features comparing with an $e^+e^-$ machine
running at the $J/\psi$ mass. 
$J/\psi$ mesons from the decay of spinless $B$ mesons are fully longitudinally polarized. This provides a useful cross check for previous measurements with
transversely polarized $J/\psi$ mesons. The charmonia from $B$ decays 
do not suffer from the beam hole effect, such that events
with $|\cos\theta|$ near 1 can be detected. 
These events are effective to determine $\alpha$. A $B$-factory is also immune from 
$e^+e^- \to q\bar{q} \to p \bar{p}$
background, where $q$ stands for a $u$ or $d$ quark. 
For previous studies this background 
is intrinsically embedded and hard to separate on an event-by-event 
basis.

We use a  357 fb$^{-1}$  data sample,
consisting of  $ 386 \times 10^6$ $B\bar{B}$ pairs,
collected by the Belle detector 
at the KEKB asymmetric energy $e^+e^-$ (3.5 on 8~GeV) collider~\cite{KEKB}.
The Belle detector is a large solid angle magnetic spectrometer
that consists of a three layer silicon vertex detector (SVD), a 50
layer central drift chamber (CDC), an array of aerogel threshold
\v{C}erenkov counters (ACC), a barrel-like arrangement of time of
flight scintillation counters (TOF), and an electromagnetic
calorimeter comprised of CsI (Tl) crystals located inside a
superconducting solenoid coil that provides a 1.5~T magnetic
field.  An iron flux return located outside of the coil is
instrumented to detect $K_L^0$ mesons and to identify muons. The
detector is described in detail elsewhere~\cite{Belle}.

In this study of two-body baryonic decays of charmonia we focus on the decay processes, $B^+ \to \ppk$
and $B^+ \to \LLK$~\cite{conjugate}. 
The event selection criteria are based on information obtained
from the tracking system
(SVD+CDC) and the hadron identification system (CDC+ACC+TOF).
All primary charged tracks
are required to satisfy track quality criteria
based on the track impact parameters relative to the
interaction point (IP). 
The deviations from the IP position are required to be within
$\pm$1 cm in the transverse ($x$--$y$) plane, and within $\pm$3 cm
in the $z$ direction, where the $z$ axis is defined by the
positron beam line. Proton, kaon and pion candidates are selected 
using $p/K/\pi$ likelihood functions obtained from the hadron
identification system. For the primary protons from $B$ decays we 
require $L_p/(L_p+L_K)> 0.6 $ and
$L_p/(L_p+L_{\pi})> 0.6$, where $L_{p/K/\pi}$ stands for the
proton/kaon/pion likelihood.
We require $L_K/(L_K+L_{\pi})> 0.3$ to identify kaons.
$\Lambda$ candidates are reconstructed from decays into the $p\pi^-$ channel. 
Each candidate must have a displaced vertex and flight direction consistent with a
$\Lambda$ originating from the interaction point~\cite{LAM}.
To reduce background, 
a $L_p/(L_p+L_{\pi})> 0.6$ requirement is applied 
to the secondary proton from the $\Lambda$ decay.

To identify the reconstructed $B$ meson candidates we use the
beam energy
constrained mass, $\mb = \sqrt{E^2_{\rm beam}-p^2_B}$, and the
energy difference, $\de = E_B - E_{\rm beam}$, where $E_{\rm
beam}$ is the beam energy, and $p_B$ and $E_B$ are the momentum and
energy of the reconstructed $B$ meson in the rest frame of
the $\Upsilon({\rm 4S})$. The candidate region is
defined as 5.2~GeV/$c^2 < \mb < 5.29$~GeV/$c^2$ and 
$ -0.1$ GeV $ < \de < 0.2$ GeV.
From a GEANT based Monte Carlo (MC) simulation, the signal
is peaked in the 
region 5.27 GeV/$c^2 < \mb < 5.29$ GeV/$c^2$ and 
$|\de|< 0.05$ GeV.

The dominant background arises from the continuum $e^+e^-
\to q\bar{q}$ process.
The background from $b \to c$ and charmless mesonic decays is 
negligible.
In the $\Upsilon({\rm 4S})$ rest frame,
continuum events are jet-like while
$B\bar{B}$ events are more spherical. 
The reconstructed momenta 
of final state particles is used to form various shape variables (e.g. thrust
angle, Fox-Wolfram moments, etc.) in order to categorize each event.  
We follow the scheme defined in Ref.~\cite{etapk} that
combines seven event shape variables into
a Fisher discriminant to suppress
continuum background.

Probability density functions (PDFs) for the Fisher discriminant and
the cosine of the angle between the $B$ flight direction
and the beam direction in the $\Upsilon({\rm 4S})$ rest frame
are combined to form the signal (background)
likelihood ${\cal L}_{s (b)}$.
The signal PDFs are determined from signal MC
simulation; the background PDFs are obtained from 
the side-band data 
with $\mb < 5.26$ GeV/$c^2$.
We require
the likelihood ratio ${\cal R} = {\cal L}_s/({\cal L}_s+{\cal L}_b)$ 
to be greater than 0.4 for both
$\ppk$ and $\LLK$ modes.
These selection criteria suppress approximately 69\% (66\%) of the background while 
retaining 92\% (91\%) 
of the signal for the $\ppk$ ($\LLK$) mode.
In this study only one $B$ candidate is allowed per event. 
If there are  multiple $B$ candidates in one event, we 
select the one with the best $\chi^2$ value from the vertex fit.
Multiple $B$ candidates are found in less than 2\% (5\%) of events
for the $\ppk$ ($\LLK$) mode. 

\begin{figure}[htb]
\epsfig{file=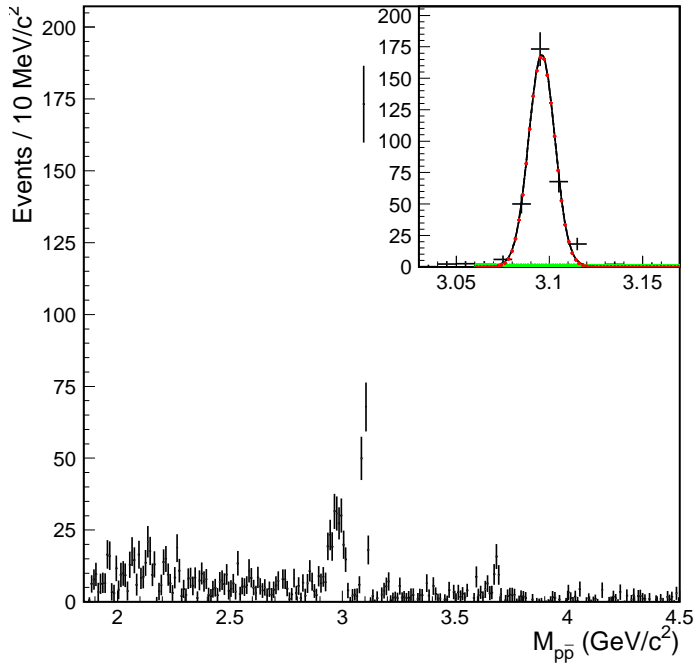,width=3.9in}
\caption{$B$ yield versus 
$M_{\rm p\bar p }$. The inset shows the $J/\psi$ mass region.(green dots,
red dots, and solid line represent fitting background shape, signal shape, and combined
result, respectively)}
\label{fg:mpp}
\end{figure}

We use an unbinned likelihood fit 
to estimate the $B$ yield:
$$ L = {e^{-(N_s+N_b)} \over N!}\prod_{i=1}^{N} 
[N_s P_s(M_{{\rm bc}_i},\Delta{E}_i)+
N_b P_b(M_{{\rm bc}_i},\Delta{E}_i)],$$
where $P_s(P_b)$ denotes the signal (background) PDF, 
$N$ is the number of events in the fit, and $N_s$ and $N_b$
are fit parameters.
For the signal PDF,
we use a Gaussian in $\mb$ and a double Gaussian in $\de$.  We fix
the parameters of these functions to values determined by MC simulation~\cite{correction}.
Background shapes are fixed from fitting to sideband
events in the region:
3.14 GeV/$c^2$ $ < M_\pp <$ 3.34 GeV/$c^2$.
The $\mb$ background is modelled using a parametrization first used by the ARGUS collaboration,
$ f(\mb)\propto \mb\sqrt{1-x^2}
\exp[-\xi (1-x^2)]$,  
where $x$ is defined as $\mb/E_{\rm beam}$ and $\xi$ is
a fixed value. 
The $\de$ background shape is modeled by a first order polynomial.


As the mass resolution of $M_\pp$ ($M_\LL$) is about 10 MeV/$c^2$, we
determine the $B$ yield as a function of $M_\pp$ ($M_\LL$) from 1.85 GeV/$c^2$ to
4.5 GeV/$c^2$ in 10 MeV/$c^2$ bins. The result is shown in
Fig.~\ref{fg:mpp} (Fig.~\ref{fg:mll}).
There are clear $\eta_c$ and $J/\psi$ peaks in the mass spectrum.
A fit to the data is shown in the inset.
We use 
a Breit-Wigner function for the $\eta_c$ peak, a Gaussian for the $J/\psi$ peak and 
a line for the non-charmonium background. 
The background
is negligible. We define the 
$\eta_c$ signal region as 2.94 GeV/$c^2 < M_\pp < 3.02 $ GeV/$c^2$ and
the $J/\psi$ signal region as  3.06~GeV/$c^2 < M_\pp < 3.14$~GeV/$c^2$.
The measured B yield is $329 \pm 19$ $(195 \pm 15)$ for $B^+ \to
J/\psi K^+, J/\psi \to \pp$ $(B^+ \to \eta_c K^+, \eta_c \to \pp)$.
We use a phase space MC sample to determine the efficiency in this $\mpp$ range.
The obtained efficiency is $38\% $ $(36\% )$ for $B^+ \to
J/\psi K^+, J/\psi \to \pp$ $(B^+ \to \eta_c K^+, \eta_c \to \pp)$.
The measured branching fractions for charmonia decaying into $\pp$ are
${\mathcal B}(\eta_c \to \pp) = ( 1.58 \pm 0.12 (stat)  \pm 0.22 (syst) \pm 
0.47 ({\rm PDG}) )
\times 10^{-3}$ and 
${\mathcal B}(J/\psi \to \pp) = ( 2.24 \pm 0.13 (stat) \pm 0.31 (syst) 
\pm 0.01 ({\rm PDG}))
\times 10^{-3}$, where the last errors are related to the uncertainty in the current world average values of the branching ratios 
${\mathcal B}(B^+ \to \eta_c K^+)$ 
and ${\mathcal B}(B^+ \to J/\psi K^+)$.

\begin{figure}[htb]
\epsfig{file=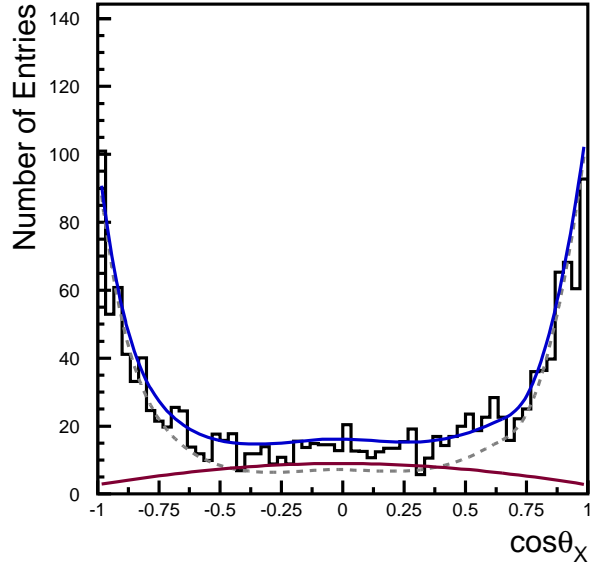,width=3.5in}
\caption{Likelihood fit of the $J/\psi\to p\bar p$ helicity angle
distribution. The blue solid, red solid, and dashed line represent the fit
results, the signal shape, and the background shape, respectively.}
\label{fg:jpsipp}
\end{figure}


\begin{figure}[htb]
\epsfig{file=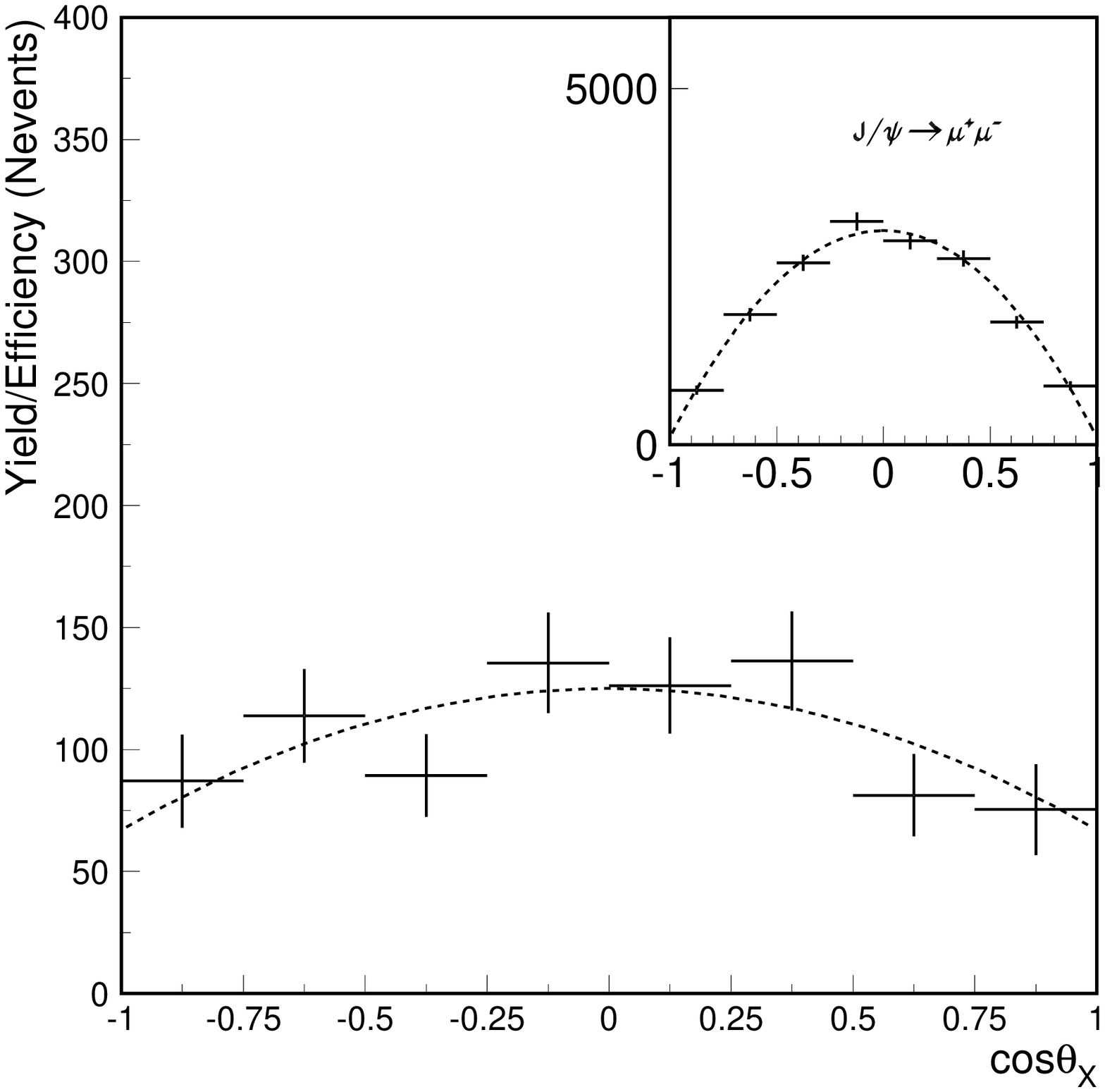,width=3.9in}
\caption{$J/\psi\to p\bar p$
helicity angle distribution. The dashed line
shows the $\chi^2$ fit result for $B$ events of
$B^+\to J/\psi K^+, J/\psi\to p\bar p$. 
The inset shows the $\chi^2$ fit result for
$B$ yield of $B^+\to J/\psi K^+, J/\psi \to \mu^+\mu^-$.}
\label{fg:comb-cos}
\end{figure}

We study the proton angular distribution 
in the helicity frame of the $J/\psi$. $\theta_X$ is defined as the
angle between the proton flight direction and the direction opposite
to the flight of the kaon in the $J/\psi$ rest frame.
The angular distribution is parameterized as $ P(\cos\theta_X) = 
(1+\alpha_B \cos^2\theta_X)/(2+2/3\alpha_B)$. 
Note that $\alpha_B$ determined from
longitudinally polarized $J/\psi$ is related to the $\alpha$
determined from transversely polarized $J/\psi$
by $\alpha_B = {-2 \alpha \over (\alpha + 1)}$~\cite{helform}.
From previous measurements of $\alpha$, the expectation of $\alpha_B$ is
$-0.80 \pm 0.04$. 
We modify the likelihood function to 
$$ L = {e^{-(N_s+N_b)} \over N!}\prod_{i=1}^{N} 
[N_s P_s(M_{{\rm bc}_i},\Delta{E}_i)
 \epsilon(\cos\theta_X) P(\cos\theta_X)+
N_b P_b(M_{{\rm bc}_i},\Delta{E}_i,\cos\theta_X)],$$
where $\epsilon(\cos\theta_X)$ is the normalized efficiency function.
The observed distribution of $\epsilon(\cos\theta_X)$ is flat. We assume there is no correlation between $\mb$, $\de$ and
$\theta_X$ based on a study with MC data. 
The background PDF, including $\cos\theta_X$, is determined from
$M_\pp$ sideband data. The distribution of $\cos\theta_X$ for
$J/\psi$ candidates and fit result in whole $\mb,\de$ region is shown in Fig.~\ref{fg:jpsipp}. $\alpha_B$ is determined to 
be $-0.54 \pm 0.14$.

As a cross check, we fit 
the $1+\alpha_B\cos^2\theta_X$ parametrization to the efficiency corrected $B$ yield 
as a function of $\cos\theta_X$. The results of the fit are
shown in Fig.~\ref{fg:comb-cos}. The
value of $\alpha_B$ obtained from the fit is $ -0.46 \pm 0.16$ with $\chi^2/d.o.f. = 1.1$. 
This is consistent
with the likelihood result from a toy MC study, where samples
with the same number of events as our data 
are generated to check the $\alpha_B$ difference between the likelihood and $\chi^2$ methods.
The $\alpha_B$ difference follows a Gaussian
distribution with a width of approximately $0.05$. We also apply both the likelihood and $\chi^2$ methods
to a $J/\psi \to \mu^+\mu^-$ event sample. The result is shown
as an inset in Fig.~\ref{fg:comb-cos}. It is in excellent agreement
with theoretical prediction which has a $\sin^2\theta_X$ shape ($\alpha_B
= -0.999 \pm 0.003$).

We study the systematic error of $\alpha_B$ by varying the value of 
various selection cuts and parameters of PDFs to check for trends in
the value of $\alpha_B$. This relation is smooth
and can be fitted to a line. 
We then quote the change in $\alpha_B$ along the line between 
the selected point
and the far end of the tested region as a systematic error. Note that
this is a conservative estimation, since the statistical fluctuation 
of this data set also contributes to changes in $\alpha_B$. 
We assign a systematic error of $0.08$ for the ${\cal R}$ selection,
$0.06$ for PID selection, and $0.02$ for fitting PDFs. Other systematic errors are
negligible. To be conservative, we also quote the observed  
difference between the likelihood method and the $\chi^2$ method as a
systematic error. The total systematic uncertainty in $\alpha_B$ is 0.13.

\begin{figure}[htb]
\epsfig{file=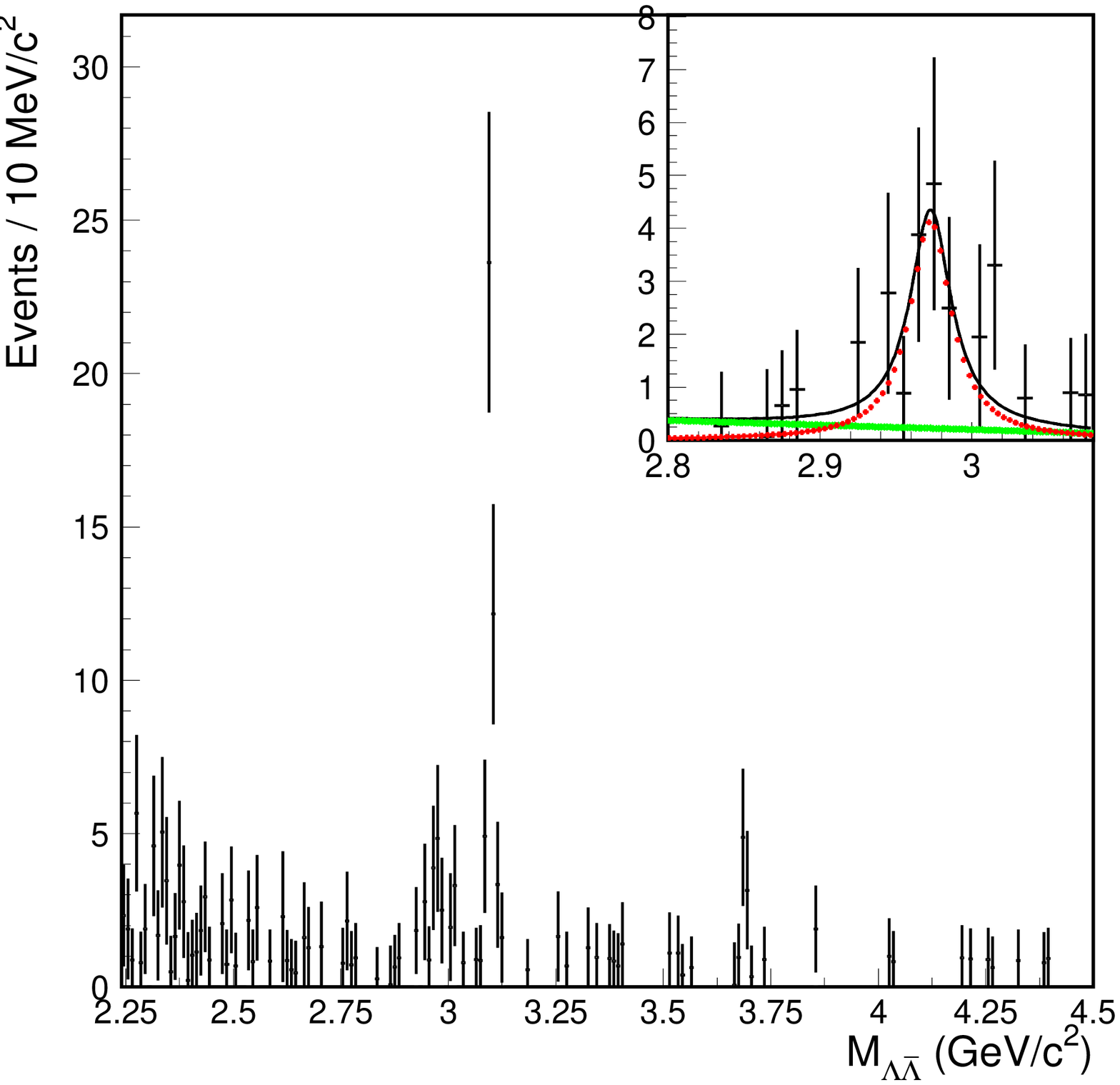,width=3.9in}
\caption{$B$ yield versus $M_{\rm\Lambda\bar\Lambda}$. The inset shows the
$\eta_{\rm c}$ mass region.(green dots, red dots, and solid line represent
fitting background shape,
signal shape, and combined result, respectively)}
\label{fg:mll}
\end{figure}

The baryon-antibaryon mass spectrum from
$B^+ \to \LLK$ decays is shown in Fig.~\ref{fg:mll}.
Similar structures are seen in the mass spectrum of $B^+ \to \LLK$ decays as were seen in the $B^+ \to \ppk$ mass spectrum.
There are several complicating factors in the analysis of $B^+ \to \LLK$ decays, relative to $B^+ \to \ppk$ decays.
The slow pion from $\Lambda$ decays has a low detection efficiency.
This causes
the $\Lambda$ reconstruction efficiency to be non-uniform in the polar angle
($\theta_p$) of the secondary decay proton in the $\Lambda$ helicity frame,
and is correlated with $\cos\theta_X$, where $X$ refers to the $\Lambda$.
The likelihood function becomes
$$ L = {e^{-(N_s+N_b)} \over N!}\prod_{i=1}^{N}
[N_s P_s(M_{{\rm bc}_i},\Delta{E}_i)
 \epsilon(\cos\theta_X,\cos\theta_p,\cos\theta_{\bar p})
P(\cos\theta_X,\cos\theta_p,\cos\theta_{\bar p} )$$ $$
+N_b P_b(M_{{\rm bc}_i},\Delta{E}_i,\cos\theta_X,\cos\theta_p,
\cos\theta_{\bar p})],$$
where $\epsilon(\cos\theta_X,\cos\theta_p,\cos\theta_{\bar p})$ 
is determined by a huge signal MC sample with $ 4 \times 10^6$ events.
The background PDF is determined from
$M_\LL$ sideband data in the region 3.14 GeV/$c^2$ $ < M_\LL <$ 3.54 GeV/$c^2$.
The value of $\alpha_B$ obtained from the fit is $-0.63 \pm 0.46 \pm 0.27$, where
the systematic error is determined from the same procedure used for $J/\psi \to \pp$ decays.

The $\mb$ distribution (with $|\de|<$ 0.05
GeV) and the $\de$ distribution (with $\mb >$ 5.27 GeV/$c^2$)
for $B^+ \to \eta_c K^+, \eta_c \to \LL$ decays are
shown in Fig.~\ref{fg:etac_2dp}.
$\eta_c$ signal peaks are visible in the $\mb$ and $\de$ distributions.
The yield from the fit is $ 19.5^{+5.1}_{-4.4}$ 
with a
statistical significance of $8.1$ standard deviations. The significance is
defined as $\sqrt{-2{\rm ln}(L_0/L_{\rm max})}$, where $L_0$ and
$L_{\rm max}$ are the likelihood values returned by the fit with
signal yield fixed at zero and its best fit value,
respectively. The fit yield is consistent with that obtained 
in Fig.~\ref{fg:mll}. We estimate the branching fraction from the ratio
of the efficiency corrected yield of $\eta_c \to \LL$ and $\eta_c \to \pp$.
The result is
${\mathcal B}(\eta_c \to \LL) = ( 0.87^{+0.24}_{-0.21} (stat) \pm 0.14 (syst)
\pm 0.27 ({\rm PDG}))
\times 10^{-3}$, where the last error is associated with the world average value for ${\mathcal B}(\eta_c \to \pp)$
. We apply the same procedure to obtain
${\mathcal B}(J/\psi \to \LL) = ( 2.00^{+0.33}_{-0.29} (stat) \pm 0.34 (syst)
\pm 0.08 ({\rm PDG}))
\times 10^{-3}$. 

\begin{figure}[htb]
\epsfig{file=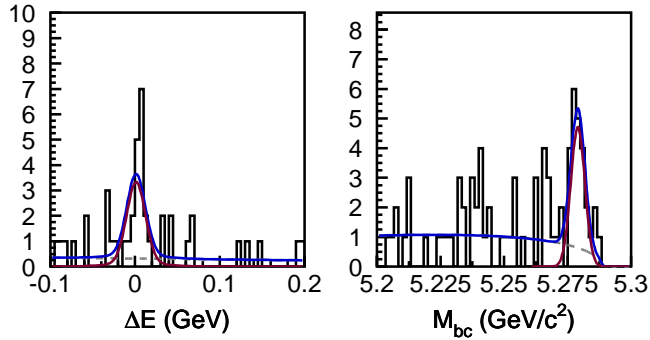,width=3.8in}
\caption{$\de$ and $\mb$ distribution of
$B^+\to\eta_{\rm c} K^+,\eta_{\rm c}\to\Lambda\bar\Lambda$
candidates. The blue solid, red solid, and dashed line represent the fit
results, the signal shape, and the background shape, respectively.}
\label{fg:etac_2dp}
\end{figure}

Systematic uncertainties 
are studied using high statistics control samples. For proton
identification, we use a  $\Lambda \to p \pi^-$ sample, while for
$K/\pi$ identification we use a $D^{*+} \to D^0\pi^+$,
 $D^0 \to K^-\pi^+$ sample.
Tracking efficiency is studied with
fully and partially reconstructed $D^*$ samples.
The $\cal R$ continuum suppression uncertainty is studied with 
$b \to c$ control samples with similar final states.
For $\Lambda$ reconstruction, we have an additional uncertainty on the
efficiency for detecting tracks away from the IP. The size of this uncertainty is determined from the
difference between $\Lambda$ proper time distributions in data and MC
simulation. 
Based on these studies,
we assign a 1\% error for each track, 2\% for each proton identification,
1\% for each kaon/pion identification, an additional 3\% for $\Lambda$
reconstruction and 3\% for the $\cal R$ selection.

The systematic uncertainty  in the fit yield is studied by varying
the parameters of the signal and background PDFs and is approximately
5\%. The MC
statistical uncertainty and modeling 
contributes a 5\% error. 
The error on the
number of $B\bar{B}$ pairs is determined to be 1\%, where
the branching fractions of $\Upsilon({\rm 4S})$ 
to neutral and charged $B\bar{B}$ pairs are assumed to be equal. 
Although the background in the $\mpp$ and $M_\LL$
spectra appear negligible, we forced the $B$ yield to be positive and re-fit the spectra.
The feed-down background is estimated to be 8\% and 12\% for
the $\pp$ and $\LL$ modes, respectively.

To produce a combined systematic error, the correlated errors are added linearly and then combined with the
uncorrelated errors in quadrature. The total systematic
uncertainties are 14\% and 17\% for
the $\ppk$, and $\LLK$ modes,
respectively.

In summary, using  $386 \times 10^6$ $B\bar{B}$ events, we measure the
branching fractions of $J/\psi \to \pp$, $\eta_c \to \pp$, $J/\psi \to
\LL$ and $\eta_c \to \LL$ from $B^+ \to \ppk$ and $B^+ \to \LLK$ decays. 
We report the first observation of $\eta_c \to \LL$ decays, with
${\mathcal B}(\eta_c \to \LL) = ( 0.87^{+0.24}_{-0.21} \pm 0.14 \pm 0.27)
\times 10^{-3}$.
We also measure the parameter $\alpha_B$ for baryonic $J/\psi$ decays.
The measured values are $-0.54 \pm 0.14 \pm 0.13$ and $ -0.63 \pm  0.46 \pm 
0.27$ for 
$J/\psi \to \pp$ and $J/\psi \to \LL$, respectively. 
The above measurements  are in agreement
with the current world average values as shown in TABLE~\ref{alphalist}. The $B$-factories will rapidly
accumulate charmonia decays in the coming years, enabling more accurate
cross checks.

\begin{table}[htb]
\caption{List of $\alpha$ in previous experiments}
\label{alphalist}
\begin{center}
\begin{tabular}{|c|c|c|}
\hline
\hline
Coll./Mode&
$J/\psi \to \pp$&
$J/\psi \to \LL$
\\
\hline
Mark1&
$1.45 \pm 0.56$&

\\
Mark2&
$0.61 \pm 0.23$&
$0.72 \pm 0.36$

\\
Mark3&
$0.58 \pm 0.14$&

\\
DASP&
$1.70 \pm 1.70$&

\\
DM2&
$0.62 \pm 0.11$&
$0.62 \pm 0.22$

\\
BES&
$0.68 \pm 0.06$&
$0.52 \pm 0.35$

\\
\hline
world average&
$0.66 \pm 0.05 (\alpha_B = -0.80 \pm 0.04)$&
$0.62 \pm 0.17 (\alpha_B = -0.77 \pm 0.13)$

\\
\hline
\hline
\end{tabular}
\end{center}
\end{table}

We thank the KEKB group for the excellent operation of the
accelerator, the KEK Cryogenics group for the efficient
operation of the solenoid, and the KEK computer group and
the National Institute of Informatics for valuable computing
and Super-SINET network support. We acknowledge support from
the Ministry of Education, Culture, Sports, Science, and
Technology of Japan and the Japan Society for the Promotion
of Science; the Australian Research Council and the
Australian Department of Education, Science and Training;
the National Science Foundation of China under contract
No.~10175071; the Department of Science and Technology of
India; the BK21 program of the Ministry of Education of
Korea and the CHEP SRC program of the Korea Science and
Engineering Foundation; the Polish State Committee for
Scientific Research under contract No.~2P03B 01324; the
Ministry of Science and Technology of the Russian
Federation; the Ministry of Education, Science and Sport of
the Republic of Slovenia; the National Science Council and
the Ministry of Education of Taiwan; and the U.S.\
Department of Energy.

\clearpage


\begin{thebibliography}{99}
\bibitem{conjugate} {Throughout this report,
inclusion of charge conjugate mode is always implied unless
otherwise stated.}
\bibitem{correction}{ There are small corrections applied to these
parameters based on the 
study of the difference between data and MC of $B \to D\pi$ decays.}
\end{thebibliography}
\end{document}